\documentclass[twocolumn,showpacs,preprintnumbers,amsmath,amssymb]{revtex4}
\usepackage{graphicx}% Include figure files
\usepackage{dcolumn}% Align table columns on decimal point
\usepackage{bm}% bold math

\begin{document}

%\preprint{APS/123-QED}

\title{Discrete model for laser driven etching and microstructuring of metallic surfaces}% Force line breaks with \\

\author{Alejandro Mora}
\email{ihram@ihr.uni-stuttgart.de}
\affiliation{Institut f\"ur H\"ochstleistungsrechnen (IHR), University of Stuttgart, \\Nobelstr. 19, D-70569 Stuttgart, Germany}
%\homepage{http://www.csv.ica.uni-stuttgart.de/homes/am/}

\author{Thomas Rabbow}
\email{rabbow@uni-bremen.de}
\affiliation{Institut f\"ur Angewandte und Physikalische Chemie, Chemische Synergetik,\\ University of Bremen, Bibliotheksstra\ss e NW2, D-28359 Bremen,Germany}
%\homepage{http://www-user.uni-bremen.de/~plath/thomas.html}

\author{Peter J\"org Plath}
\email{plath@uni-bremen.de}
\affiliation{Institut f\"ur Angewandte und Physikalische Chemie, Chemische Synergetik,\\ University of Bremen, Bibliotheksstra\ss e NW2, D-28359 Bremen,Germany}
%\homepage{http://www-user.uni-bremen.de/~plath/}

\author{Maria Haase}
\email{ihrmh@ihr.uni-stuttgart.de}
\affiliation{Institut f\"ur H\"ochstleistungsrechnen (IHR), University of Stuttgart, \\Nobelstr. 19, D-70569 Stuttgart, Germany}
%\homepage{http://www.csv.ica.uni-stuttgart.de/homes/mh/}

\date{\today}% It is always \today, today,
             %  but any date may be explicitly specified

\begin{abstract}
We present a unidimensional discrete solid-on-solid model evolving in time using a kinetic Monte Carlo method to simulate micro-structuring of kerfs on metallic surfaces by means of laser-induced jet-chemical etching. The precise control of the passivation layer achieved by this technique is responsible for the high resolution of the structures. However, within a certain range of experimental parameters, the microstructuring of kerfs on stainless steel surfaces with a solution of $\mathrm{H}_3\mathrm{PO}_4$ shows periodic ripples, which are considered to originate from an intrinsic dynamics. The model mimics a few of the various physical and chemical processes involved and within certain parameter ranges reproduces some morphological aspects of the structures, in particular ripple regimes. We analyze the range of values of laser beam power for the appearance of ripples in both experimental and simulated kerfs. The discrete model is an extension of one that has been used previously in the context of ion sputtering and is related to a noisy version of the Kuramoto-Sivashinsky equation used extensively in the field of pattern formation.
\end{abstract}
% DOI: 10.1103/PhysRevE.72.061604
% PACS, the Physics and Astronomy
% Classification Scheme.
% 61.82.Bg Radiation effects on metals and alloys
% 81.65.Cf Surface cleaning, etching, patterning in surface treatments
% 05.10.Ln Monte Carlo methods (statistical physics/nonlinear dynamics)
% 47.54.+r Pattern selection; pattern formation (fluid dynamics)
%\keywords{Suggested keywords}
%Use showkeys class option if keyword
%display desired

\maketitle

\section{Introduction}

Since the 1980s, laser induced wet chemical etching in silicon, ceramics, and metals has been an intensively used micro-structuring technique \cite{gutfeld:1998}. Controlling the quality of the final structures is the main issue from the experimental and theoretical point of view. In a variation of the experimental technique developed by Metev, Stephen and collaborators (see Refs. \cite{nowak:1996} and \cite{stephen:2004}) and currently implemented by Rabbow et al. (see Ref. \cite{rabbow:2005}), a focused laser beam induces an etching reaction enhanced by a coaxial jet of etchant producing holes and kerfs on metallic samples within a micrometer scale. 

This technique is called \textit{laser induced jet-chemical etching} (LJE) and has been successful in fabricating superelastic microgrippers of nickel-titanium alloy. However, within a certain range of parameters, in the microstructuring of kerfs on stainless steel surfaces with a solution of $\mathrm{H}_3\mathrm{PO}_4$, unwanted periodic ripples appear. In this paper, it is considered that these ripples are intrinsically generated and belong to a wide universality class of pattern formation phenomena that emerge, for example, in ripple structures formed by wind over a sand bed, ion sputtering of various surfaces \cite{makeev:2002,valbusa:2002} and abrasive water-jet cutting \cite{friedrich:2000}.

In the context of ion sputtering, Cuerno, Makse, Tommassone, Harrington and Stanley (CMTHS) proposed a stochastic one dimensional (1D) discrete solid-on-solid (SOS) model based on the competition between erosion and surface diffusion processes in which ripples appear at early stages of the evolution \cite{cuerno:1995c}. We have extended and adapted the CMTHS model for the LJE case taking into account a moving Gaussian-distributed laser beam which leads to a localized heating and etching of the metallic sample. The motivation to use such a phenomenological model is originated from the limited knowledge of the interaction of diverse processes occurring in a wide range of scales. The dynamics of the laser light absorption, heat, chemical reactions, hydrodynamics, and transport phenomena is too complex to be fully modelled. While this \textit{extended model} is quite simple, it nevertheless captures essential physical effects of the process, such as the unstable temperature driven etching and the stabilizing mechanism of surface diffusion. In addition, since the model is evolved in time using a kinetic Monte Carlo method, fluctuations and rough surfaces are produced naturally.

This paper is organized as follows. In Sec. \ref{sec:experiment} we present the basics of the experimental setup and describe qualitatively some of the relevant microscopic processes occurring during etching. The appearance of a ripple regime in a series of kerfs structured with increasing laser power is analyzed in Sec. \ref{sec:obtained_ripples}. A hypothesis about the intrinsic nature of the ripples is proposed there. In Sec. \ref{sec:model} we describe and justify the extension of the CMTHS model analyzing the scaling properties and ripple regimes for uniform erosion. The application of the extended model to simulate the LJE is presented in Sec. \ref{sec:laser_jet_action}. Varying the simulated laser power, a ripple regime is obtained and compared with the experimental one. In the conclusions we summarize achievements and shortcomings of the model and suggest future improvements.

\begin{figure}
\begin{center}
\includegraphics[width=0.482\textwidth]{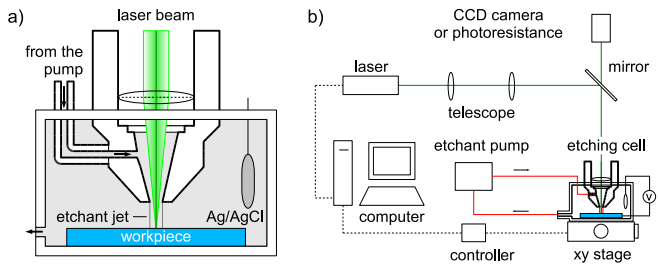}\caption{\label{fig:exp_setup}(Color online) Schematic diagram of the experimental setup. (a) The etching cell. A focused laser induces an etching reaction enhanced by a coaxial running jet of etchant.
(b) The whole experimental setup. The reaction can be observed with a charged-coupled device (CCD) camera and the intensity of the reflected light is measured with a photoresistance. Due to the laser induced etching the workpiece changes its state locally from passive to active. Changes of the potential are detected with an Ag/AgCl reference electrode.}
\end{center}
\end{figure}

\section{LJE technique}\label{sec:experiment}

\subsection{Experimental setup}

A schematic diagram of the etching cell is shown in Fig. \ref{fig:exp_setup}(a). In this implementation of the technique, foils of stainless steel (Fe/Cr18/Ni10) are immersed horizontally in a solution of etchant based on 5-M $\mathrm{H}_3\mathrm{PO}_4$. The microstructuring of the samples is achieved by an argon ion cw-laser beam at 514nm embedded coaxially in a jet of liquid etchant which is directed perpendicularly to the surface.
The intensity profile of the focused beam is assumed to be Gaussian with an estimated standard deviation of $\sigma\approx2\ \mu$m. The laser spot diameter for the experiment is defined as $d_{exp}\equiv4\sigma\approx8\ \mu$m. The laser induced etching leads to dissolution of the metal and formation of hydrogen. Simultaneously, changes of the state of the electrode from passive to active produce an electrochemical potential and its temporal evolution $E(t)$ can be measured against an Ag/AgCl reference electrode, which is immersed in the etching reservoir. 

The etching cell is mounted on a computer-controlled mobile basis which allows us to move the sample in the $xy$-plane with respect to the laser beam with different feed velocities $v_f$ [see Fig. \ref{fig:exp_setup}(b)]. The setup is automated allowing us to structure holes or kerfs varying the most relevant external parameters: laser power, etchant jet velocity, feed velocity, and etchant concentration. Details of the process can be observed with a CCD camera located above the etching cell in the same axis of the laser beam. Alternatively, by means of a photoresistance located in the same position, measurements of the intensity of the reflected light can be used together with the electrochemical potential for monitoring the etching dynamics.

\subsection{Microscopic description of the etching process}\label{sec:microscopic_process}

Under normal conditions of temperature, the layer in contact with the liquid is passivated spontaneously thus isolating the metallic sample from the etchant action. The area below the laser spot is heated and almost immediately the heat spreads to a wider zone. Above a temperature threshold the passivation layer is removed and thermally activated chemical etching starts there forming the \textit{etching front} which structures the kerf on the surface at velocity $v_f$. The protons of the phosphoric acid ($\mathrm{H}_3\mathrm{PO}_4$) react with the iron, nickel, and chromium of the steel producing hydrogen and dissolution of metal ions. The basic reactions can be described as
\begin{equation}
\mathrm{Reduction\ (formation\ of\ hydrogen):\ }2\mathrm{H}^++2\mathrm{e}^-   \rightarrow \mathrm{H}_2\uparrow
\end{equation}
\begin{equation}
\mathrm{Oxidation\ (ionization\ of\ the\ metal):\ } \mathrm{Fe} \rightarrow \mathrm{Fe}^{2+}+2\mathrm{e}^{-}
\end{equation}
with similar reactions for the ionization of the nickel and chromium. The etching reaction is exothermic in nature.

When the etching reaction dissolves the metal and consumes the etchant, a thin layer of solution in contact with the metallic surface develops a concentration gradient ranging from zero on the surface to the value of the bulk concentration. This is called the \textit{Nernst diffusion layer} (NDL) and within its thickness $\delta$, the transport of ions of etchant and reaction products of the reactions occurs exclusively by diffusion, limiting the etching reaction \cite{bard:electrochemical}. Outside this layer, convective transport maintains the concentration uniform at the bulk concentration. The value of $\delta$ depends on the hydrodynamic conditions imposed by the etchant jet. 

The NDL is not directly modelled in this work. We assume that its thickness $\delta$ is almost negligible compared with the dimensions of the simulated topographies. It is worthwhile to note that the role of the Nernst diffusion layer is indeed essential in the real etching process. Variations of the dynamics of the jet due to its interaction with the surface can affect the value of the thickness $\delta$ and in consequence, the transport of ions and reaction products. For example, if in a trough of a ripple the turbulence of the etchant flow allows a growing thickness $\delta$, this would imply a strong inhibition of the etching rate after a certain depth threshold. 

In experiments of wet-chemical etching of ceramics (without etchant jet), Lu \textit{et al.} have found that the diffusive transport in the NDL is the main limiting factor of the etching rate \cite{lu:1988}. Following their analysis and assuming that the etching rate decreases exponentially with the depth, an expression for the dependency of the etched depth with the feed velocity has been found to be in good agreement with experimental data for the LJE experiment using $5-\mathrm{M}\ \mathrm{H}_3\mathrm{PO_4}$ (see Ref. \cite{rabbow:2005}). 
The function of the etchant jet is to enhance the etching rates reducing the NDL thickness $\delta$, to provide fresh etchant and to remove dissolved material and reaction products. In addition, the jet creates a cooling effect that maintains the heating effect of the laser spot concentrated in a small region. Therefore the de-passivated zone and the resulting etching are highly localized. When the etchant jet velocity is increased the etching reaction is favored due to more fresh etchant but on the other hand it is inhibited due to the cooling effect. After the laser jet leaves a region the surface is repassivated due to the decrease of the temperature and its topography will remain unaltered.  

\section{Obtained kerfs and intrinsic ripple formation}\label{sec:obtained_ripples}

\begin{figure}
\begin{center}
\includegraphics[width=0.482\textwidth]{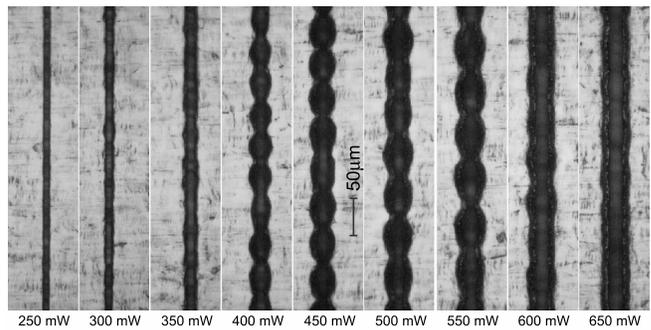}\caption{\label{fig:power_series}Reproduced from Rabbow \cite{rabbow:2005}. Optical microscope images showing a series of kerfs structured with increasing powers from 250 to 650 mW (etchant 5-M $\mathrm{H}_3\mathrm{PO_4}$, feed velocity $6\ \mu$m/s, etchant jet velocity 190 cm/s). The kerfs for 250, 300, and 350 mW are the result of a uniform etching front which widens with power. For powers greater than 350 mW, an instability in the etching front comes in, and the obtained ripples seem to be product of a thermal runaway. A hypothesis on the intrinsic character of the ripples and some remarks on the mechanism that is creating this pulsating etching front are discussed in the text.}
\end{center}
\end{figure}

In most cases the etching front is stationary and the obtained kerf has, apart from small fluctuations, a defined shape with almost constant width and depth. The bottom and walls of the kerfs are rough due to the stochastic character of the etching reactions. On the other hand, for certain parameters ranges of feed velocity, laser beam power, or etchant jet velocity, oscillations of the etching front producing periodic ripples have been reported \cite{rabbow:2005}. Here we will analyze the series of experiments with increasing power and a constant feed velocity $v_f=6\ \mu$m/s shown in Fig. \ref{fig:power_series}.

There are two external sources which produce periodic structures that can be detected in the measurements for stationary etching fronts. First, the peristaltic pump introduces vibrations in the etchant jet in the range of 1-3Hz that can be detected in the power spectra of the electrochemical potential time series, which correspond to periodic surface structures smaller than $10\ \mu$m. Second, the $xy$ stage has an internal mechanism that controls its position every $40\ \mu$m, resulting in small oscillations in the feed velocity $v_f$ around a mean value.  Nevertheless, the frequencies corresponding to the examined ripples are incommensurable with the frequencies of both the etchant pump and the controlling mechanism of the $xy$ stage. In consequence these can be discarded as external triggers of the obtained ripples. 

From the point of view of the theory of pattern formation in continuous media, the appearance of ripples originated from instabilities is not unexpected because the etching front is formed in a system far from equilibrium due to the continuous and combined action of the laser beam and the etchant jet. In general, one class of mechanisms for such instabilities arises from the existence of constraints and conservation laws in the system \cite{cross:1993}. In the case of water jet cutting, at high feed rates intrinsically generated ripples and striation patterns of the \textit{order} of the water jet diameter are formed at the side walls of the cut (see Ref. \cite{radons:2004}). This fact has been used there as a criterion to distinguish between ripples triggered externally from the intrinsic ones. 

The diameter of the etching front is not always constant for the LJE experiment.  Certainly, more delivered power means higher temperatures and, although the laser spot diameter remains the same, the etching front become wider. For laser powers 250, 300, and 350 mW an approximately stationary etching front produces kerfs with increasing width and depth (see first three kerfs in Fig. \ref{fig:power_series}).

For powers larger than 350 mW an instability appears producing ripples with increasing length and width. The ripples look like a product of thermal runaways which reach much broader areas than in the stationary etching front case. Each thermal runaway seems to appear above a temperature threshold, producing a quick broadening of the etching front. The broadening stops when the accelerated consumption of etchant inhibits further etching rates and the etching front shrinks until the conditions for the onset of the next thermal runaway are fulfilled. Note that during the ripple formation the absorption of laser radiation is also changing according to variations of the slope of the etching front. Probably this has a reinforcing effect on the pulsating etching front.

Based on these facts we formulate the hypothesis that above a power threshold an instability creates ripples. The resulting wavelength is in the order of magnitude of the diameter of the stationary etching front that would appear if the mechanism that causes the instability would not act. This could explain why the ripple length increases with the power. For powers larger than 600 mW the ripples seem to overlap as the kerfs become deeper.

\section{Discrete model}\label{sec:model}

Instead of formulating a microscopical model with the interaction of all possible chemical and physical processes, we propose a minimal phenomenological description of the etching process based on some ideas of the theory of far from equilibrium evolving surfaces. A number of discrete growth models and continuum stochastic equations have been proposed to describe the kinetic roughening properties of surface growth and erosion \cite{barabasi:fractal}. For simulating the structures produced by the LJE technique, we have extended and adapted a stochastic 1D discrete solid-on-solid (SOS) model proposed by Cuerno \textit{et al.} \cite{cuerno:1995c} for the evolution of ion sputtered surfaces (CMTHS model). Within that framework they explain two stages: an early time regime characterized by ripples and late time regime where the roughening shows self-affine scaling behavior. It has been shown that this model is related with a noisy version of the Kuramoto-Sivashinsky equation, which has been used extensively in the theory of unstable pattern formation \cite{lauritsen:1996}. 

\begin{figure}
\begin{center}
\includegraphics[width=0.482\textwidth]{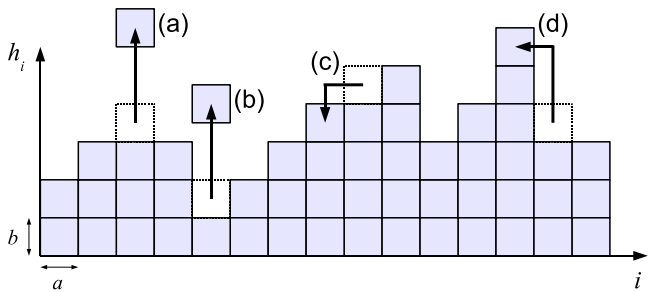}\caption{\label{fig:rules}Diagram representing the surface and the erosive and diffusive actions.  (a), (b): sites submitted to the erosion rule. The probability of it being eroded is larger for the \textquotedblleft valley" in (b) than for the \textquotedblleft peak" in (a) according to an estimation of the curvature described in Sec. \ref{sec:extended_erosion_rule}. (c), (d): diffusive movements. The probability of surface diffusion in (c) is close to 1, while for (d) it is very small according to a mechanism for creating a positive surface tension described in Sec. \ref{sec:diffusion_rule}.}
\end{center}
\end{figure}

A remarkable feature in experiments of ion sputtering is the presence of nearly periodic ripples, aligned parallel or perpendicular to the bombarding ion beam \cite{makeev:2002,facsko:1999}. Relating the energy of the ion beam with sputtering yield, Bradley and Harper (BH) \cite{bradley:1988} found that the dependency of the erosion rate with the local surface curvature induces an instability, which is responsible for the formation of periodic ripples with a characteristic length. Troughs are eroded faster than peaks and this effect can be considered as a \textquotedblleft negative surface tension," which competes with the smoothing mechanism of thermally activated surface diffusion (which is a positive surface tension).

Our extended model is based on similar mechanism and the justification for applying it to ripples found in LJE is based on experimental evidence that shows that troughs are preferentially etched as compared to peaks. This can be explained as consequence of differences in the absorption of laser energy and the resulting heat processes. First, when the laser beam is acting on a trough, due to the geometry the reflected rays converge to the zone above the trough and eventually can produce multiple reflections inside. This means that the trough and its neighborhood can effectively absorb more laser radiation, more heat is produced, and in consequence the etching rate is enhanced. In the case of peaks reflected rays are dispersed in all directions and there are no secondary reflections. On the other hand, considering the cooling effect of the etchant jet due to heat convection, it is easy to imagine that peaks are cooled more efficiently than troughs where eddies and even stagnation of the fluid are more probable to appear. Then, the preferential heating and etching within troughs result in further increase of the local curvature, which in turn enhances the secondary reflections and heating due to poorer convective heat transport by the etchant flow. 
     
\subsection{Extension of the CMTHS model}

The material to be eroded is represented in 1+1D by a lattice composed of cells of width $a$ and height $b$, and the surface is represented by the integer valued height $h_i$ where $i=1,\ldots,L$ (see Fig. \ref{fig:rules}). The system size $L$ is the number of cells in the horizontal direction and periodic boundary conditions apply for the height $h_i$. For each column, all the sites below the surface are occupied with cells, whereas all the sites above are empty. Overhangs are not allowed in this kind of model for simplifying both the simulation and the analytical approach. 

The temporal evolution of this virtual surface takes into account rules for representing the erosion and the surface diffusion processes. In this kind of kinetic Monte Carlo simulations each erosion or diffusion event appears with a rate that has to be guessed through the use of all the available experimental and theoretical information \cite{kratzer:2001}. The program defines a flat surface (one of the possible initial conditions), selects a site, and invokes with probability $f$ the erosion rule and with probability $1-f$ the surface diffusion rule. The process of selection of sites or rules to be applied is performed using a random number generator described in Ref. \cite{nrif}.

\begin{figure}
\begin{center}
\includegraphics[width=0.482\textwidth]{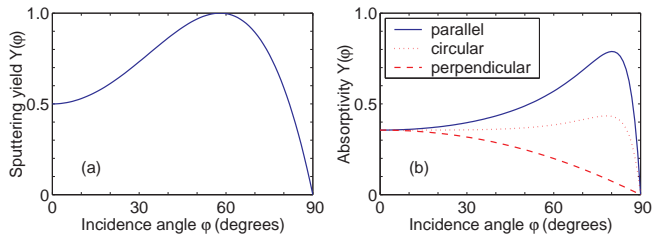}
\caption{\label{fig:absorption_combo}(Color online) (a) The sputtering yield dependence with the angle of incidence. The function used in the simulations is $Y(\varphi)=0.5+0.979\varphi^2-0.479\varphi^4$ and $Y(0^\circ)=0,Y(57.3^\circ)=1,Y(90^\circ)=0$. (b) Absorptivity of polarized light versus incidence angle for a flat iron surface (based on Ref. \cite{yao:2000}). Plane of polarization parallel to the incidence plane (continuous line), circular polarization (dotted line), plane of polarization parallel to the incidence plane (dashed line). The coefficient of refraction is $n=3.81$ and the attenuation coefficient is $k=4.44$.}
\end{center}
\end{figure}

\subsubsection{Extended erosion rule}\label{sec:extended_erosion_rule}

The erosion probability $p_e$ for a cell at the site $i$ is estimated as the product $p_e=p_\kappa Y_i$. The quantity $p_\kappa$ corresponds to the probability of being eroded depending on the curvature of the surface at the site and accounts for the unstable erosion mechanism that exists in the physical system. The value of $p_\kappa$ is larger for positive curvatures than for negative ones [see Figs. \ref{fig:rules}(a) and \ref{fig:rules}(b)]. In the CMTHS model for ion sputtering the dependency of the erosion rate on the angle of incidence $\varphi$ between the beam and a tilted portion of the surface is described by the sputtering yield function (from Ref. \cite{lauritsen:1996}):
\begin{equation}\label{eq:sputtering_yield}
Y_i=Y(\varphi_i)=y_0+y_1\varphi_i^2+y_2\varphi_i^4,
\end{equation}
where $\varphi_i$ is the incidence angle formed by the incoming beam and the normal direction defined on the surface at the site $i$. 
We can use the same function for the LJE case based of the fact that absorption of polarized light by flat metallic surfaces has similar functional dependence (in the case of electric field parallel to the plane of incidence or even circular polarization, see Fig. \ref{fig:absorption_combo}). The nonlinearity introduced by this yield (absorption) function becomes relevant at late regimes when large slopes develop, then the ripples will be strongly distorted and the surface will have a rough morphology. 

We propose to replace the \textquotedblleft box rule" described in Ref. \cite{cuerno:1995c} by a direct estimation of the angles and the curvatures based on a finite central differences method around a selected site. The first derivative or gradient of the surface at a point $i$ is $\nabla_i=(h_{i-1}-h_{i+1})/2a$ and the angle $\varphi_i$ that the surface form at this site is estimated by $\varphi_i=\arctan(\nabla_i)$. This angle corresponds to the incidence angle used in the formula (\ref{eq:sputtering_yield}). The second derivative is $\nabla^2_i=(h_{i-1}-2h_{i}+h_{i+1})/a^2$. The curvature can be estimated using the standard formula $\kappa_i=\nabla^2_i[1+(\nabla_i)^2]^{-3/2}$.

Due to the discreteness of the height $h_i$ the values obtained with these formula vary drastically from one site to the other. To attenuate this problem, the values for the angles and curvatures are computed not only for the site $i$ but also for its neighbors $i-1$ and $i+1$. Then the mean value of the angle for the site $i$ is $\overline{\varphi_i}=(\varphi_{i-1}+\varphi_i+\varphi_{i+1})/3$
and the mean curvature is: $\overline{\kappa_i}=(\kappa_{i-1}+\kappa_i+\kappa_{i+1})/3$. This procedure takes into account the values of five sites ($h_{i-2},\ h_{i-1},\ h_{i},\ h_{i+1},\ h_{i+2}$) and provides a smoothed estimation of the angles and curvatures which will influence strongly the evolution of the topography of the surface.

Taking into account these modifications in the algorithm, it is necessary to introduce two new parameters for estimating the curvature dependent erosion probability $p_\kappa$.
First, the maximum of the positive curvature $\kappa_{max}$ (the minimum $\kappa_{min}$ is the negative of this value) and  second, the minimum of the erosion probability $p_{\kappa ,{min}}$. The obtained values of the curvature $\kappa_i$ are mapped by a linear transformation in such a way that for $\kappa_i=\kappa_{max}$ the curvature dependent erosion probability is $p_\kappa=p_{\kappa ,max}\equiv1$ and when $\kappa_i=\kappa_{min}$ then $p_\kappa=p_{\kappa ,{min}}$. In the case that the computed curvature $\kappa_i$ were larger than $\kappa_{max}$ the algorithm assigns $\kappa_i=\kappa_{max}$. Similarly when $\kappa_i<\kappa_{min}$ then $\kappa_i=\kappa_{min}$. 

\subsubsection{Surface diffusion rule}\label{sec:diffusion_rule}

This rule is implemented in the same way as proposed in the original CMTHS model. The probability of a diffusive movement of a selected cell $i$ is evaluated selecting at random one of the two nearest neighbors and computing the hopping probability (see Ref.  \cite{siegert:1994}):

\begin{equation}\label{eq:hopping_probability}
w_i^{\pm}=\frac{1}{1+\exp[\Delta \mathcal{H}_{i\rightarrow i\pm 1}/(k_BT)]},
\end{equation}  
where $\Delta \mathcal{H}_{i\rightarrow i\pm 1}$ is the energy difference between the final and initial states of the move, $k_B$ is the Boltzmann constant, and $T$ is the temperature. This surface energy is defined by the Hamiltonian:
\begin{equation}\label{eq:hopping_hamiltonian}
\mathcal{H}=\frac{J}{b^2}\sum_{i=1}^{L}(h_i-h_{i+1})^2,
\end{equation} 
where $J$ is a coupling constant through which the nearest neighbor sites interact and $b$ is the height of the cells. Diffusion movements that produce final states with lower surface energy are then highly preferred [see Figs. \ref{fig:rules}(c) and \ref{fig:rules}(d)]. 

\subsection{Scaling of the extended model}\label{sec:scaling_model_uniform}

As usual in the field of far from equilibrium interfaces, we are interested in the temporal behavior of the surface width defined as
\begin{equation}
W(t)=\sqrt{\frac{1}{L}\sum_{i=1}^{L}[h_i(t)-\langle h(t)\rangle]^2},
\end{equation} 
where $L$ is the system size and $\langle h(t)\rangle$ is the mean value of the all heights of the surface at the time $t$.  The width $W(t)$ can be considered as a characterization of the roughness and it is evaluated averaging over a number of different realizations of the random number seed. In order to evaluate an eventual scaling behavior $W(t)\sim t^{\beta}$ within an interval, the growth exponent $\beta$ is estimated by the method of the consecutive slopes (see Ref. \cite{barabasi:fractal}, p. 305). For all the figures and analyses of this section, the time unit $t$ is chosen to correspond to $L$ erosion or diffusion rule invocations.

\begin{figure}
\begin{center}
\includegraphics[width=0.482\textwidth]{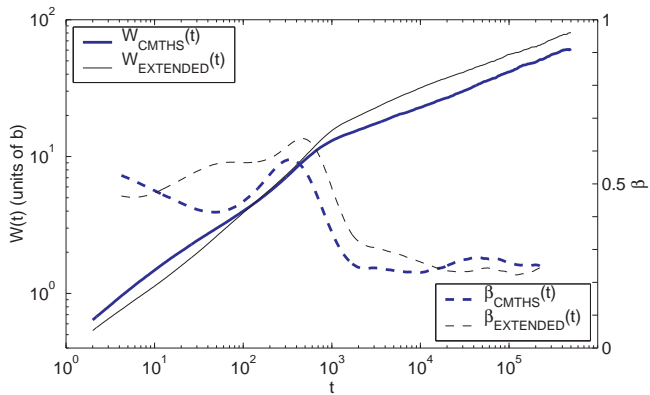}\caption{\label{fig:scaling_cuerno_closest2cuerno}(Color online) Comparison of the scaling of the surface width for the CMTHS and extended models. For the CMTHS model: The surface width $W_{\mathrm{c}}(t)$ (bold continuous line, scale on the left) and its growth exponent $\beta_{\mathrm{c}}$ (bold dashed line, scale on the right). Parameters: $L=2048$, $f=0.5$, $J/(k_BT)=5$. For the extended model: The width $W_{\mathrm{e}}(t)$ (thin continuous line, scale on the left) and its growth exponent $\beta_{\mathrm{e}}$ (thin dashed line, scale on the right). The extra parameters of the extended model are the closest possible to the CMTHS model: $a=1$, $b=1$, $p_{\kappa,min}=1/7$, $\kappa_{max}=2$.}
\end{center}
\end{figure}

In order to verify the connection of the extended model with the original CMTHS model, the evolution of the surface width for both is compared in Fig. \ref{fig:scaling_cuerno_closest2cuerno}. The system size is $L=2048$ and the width values were averaged over 100 different realizations. The probability of invoking the erosion rule is $f=0.5$ (the diffusion rule is invoked with probability $1-f=0.5$) and the constant associated with the diffusion is $J/(k_BT)=5$. The extra parameters of the extended model ($a=1$, $b=1$, $p_{\kappa,min}=1/7$, $\kappa_{max}=2$) have been chosen to reproduce as closely as possible the box rule for the erosion probability of the CMTHS model. For both cases the same \textquotedblleft yield" function is used $Y(\varphi)=0.5+0.979\varphi^2-0.479\varphi^4$.

The scaling properties are similar, showing first a rough interface at early times, then a strong increase of the growth exponent $\beta$ due to the onset of the instability that creates ripples, and finally a drop due to the stabilizing and roughening effect of the non-linear terms.  However, the scaling behavior is not identical. A precise identification of the limits of the ripple regimes is difficult because they depend on each realization and the criterion to distinguish between fluctuations and proper ripples. The ripple regime for the CMTHS model can be estimated to be $t\approx(300,1000)$; while for the extended model, the onset of the instability occurs earlier and the ripple regime lasts longer. The ripples in the extended model are larger in amplitude and present a quasi sinusoidal shape.

\begin{figure}
\begin{center}
\includegraphics[width=0.482\textwidth]{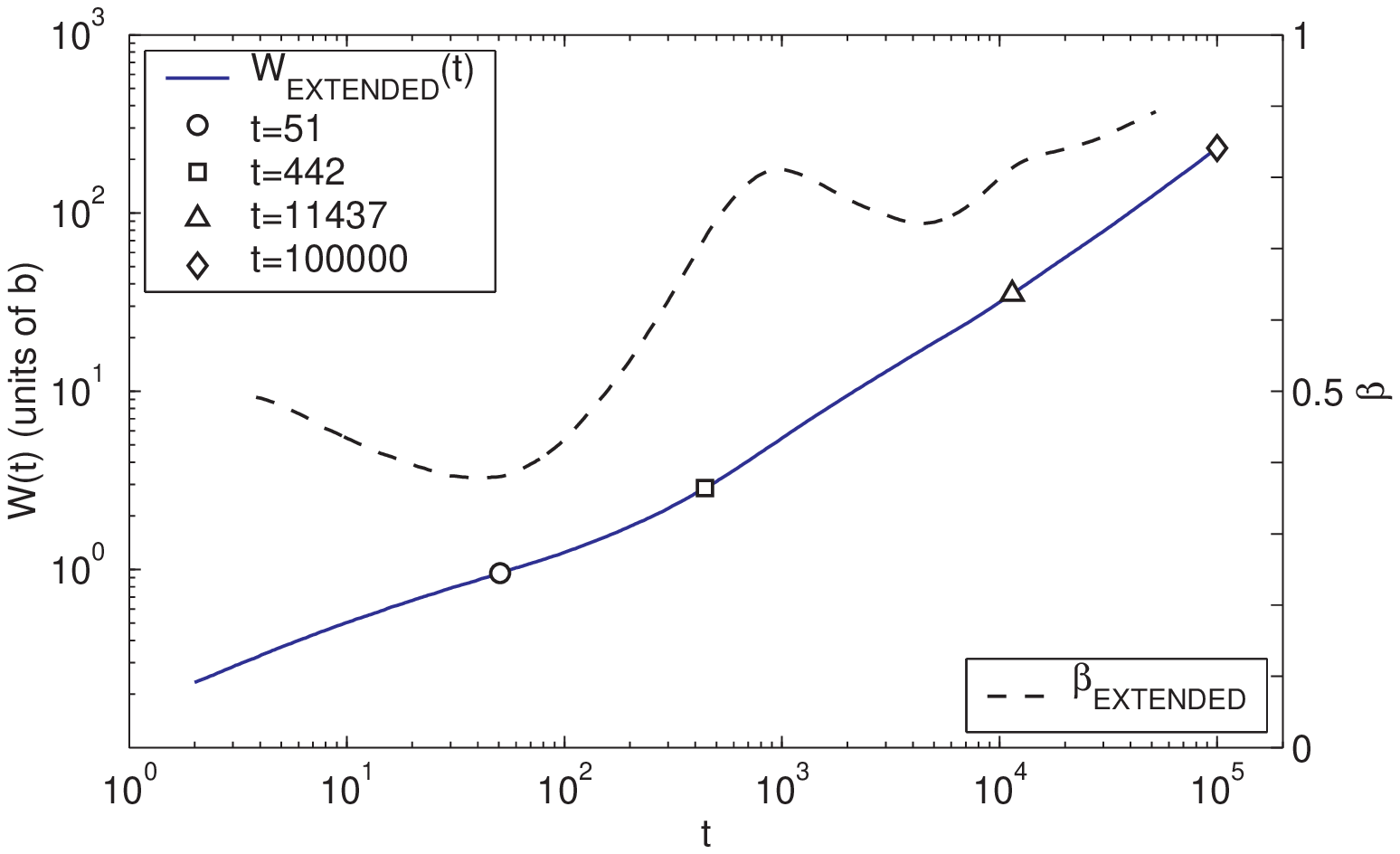}\caption{\label{fig:scaling_ripples_quasi_linear}(Color online) Temporal evolution of the interface width $W_{\mathrm{e}}(t)$ (continuous line, scale on the left) and growth exponent $\beta_{\mathrm{e}}$ (dashed line, scale on the right) of the extended model for the parameter set used in Sec. \ref{sec:laser_jet_action}. The surface's morphology at times corresponding to $t=51\ (\circ \ \mathrm{symbol})$, $t=442\ (\square \ \mathrm{symbol})$, $t=11\,437\ (\bigtriangleup \  \mathrm{symbol})$ and  $t=10^5\ (\lozenge \ \mathrm{symbol})$ is analyzed in Fig. \ref{fig:ripples_lje_stages}. Extended model parameters: $Y(\varphi)\equiv1$, $L=2048$, $f=0.045$, $J/(k_BT)=1$, $a=20$, $b=1$, $p_{\kappa ,min}=0.1$, $\kappa_{max}=0.0004$.}
\end{center}
\end{figure}
The application of the extended model for the simulation of the LJE that is presented in the next section requires to exploit the flexibility of the new parameters to generate ripples with a characteristic length in the order of the size of active etching front. In addition, the ripples should be clearly distinguishable from the inherent fluctuations and roughness produced by the stochastic character of the model. We will apply the erosion and surface diffusion rules within a moving probability distribution that is proportional to a temperature field generated by the absorption of light on the region illuminated by the focused laser. Details will be explained in Sec. \ref{sec:laser_jet_action}. The active etching zone spans a region wider than the laser spot. We will assume that variations introduced by the angle dependence of the absorption of light at the laser spot expressed by the yield (absorption) function $Y(\varphi)$ can be neglected in a first approximation.

Accordingly, in the rest of this section, we will analyze the scaling properties and the morphology evolution using $Y(\varphi)\equiv1$ and the same parameters that will be used for the simulations of the LJE presented in Sec. \ref{sec:laser_jet_action}. The probability of invoking the erosion rule is $f=0.045$ (the diffusion rule is invoked with probability $1-f=0.955$) and the constant associated with the diffusion is $J/(k_BT)=1$. For the curvature dependent erosion probability $p_\kappa$ the values $a=20$, $b=1$, $p_{\kappa,min}=0.1$, $\kappa_{max}=0.0004$ have been used. The system size is $L=2048$ and the width values were averaged over 30 different realizations. 

Figure \ref{fig:scaling_ripples_quasi_linear} shows the temporal evolution of the width $W_{\mathrm{e}}(t)$ and its growth exponent $\beta_{\mathrm{e}}$ for this parameter set. The ripple regime persists much longer than in the cases presented in Fig. \ref{fig:scaling_cuerno_closest2cuerno} and the obtained ripples have a larger amplitude. For times $t\lesssim 150$ the exponent $\beta$ is lower than the value $0.5$ characteristic for random erosion. The strong increase of the value of $\beta$ after $t\gtrsim 100$ is associated with the onset of the ripple regime. During the interval $10^3 \lesssim t\lesssim 10^4$ there is an oscillation of $\beta$ but the values remain relatively high. The origin and meaning of this oscillation is currently unknown. After $t\approx10^4$ the growth exponent $\beta$ shows a slow increase.

\begin{figure}
\begin{center}
\includegraphics[width=0.482\textwidth]{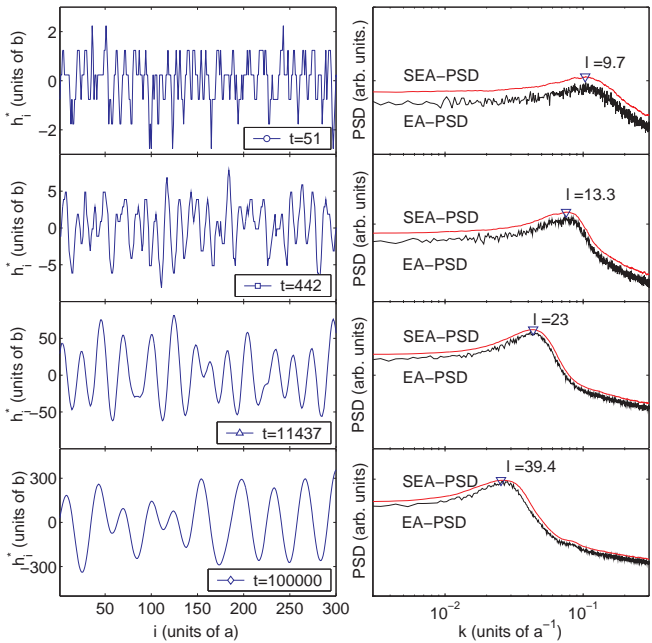}\caption{\label{fig:ripples_lje_stages}(Color online)  Four different stages of the evolution of the surface for the extended model corresponding to the times referred to in Fig. \ref{fig:scaling_ripples_quasi_linear}. The left column shows the height $h^*_i$ (note that all vertical scales are different). The real system size is $L=2048$ but only $300$ points are shown in order to appreciate details of the morphology of the surface. The right column shows log-log plots of the corresponding \textit{ensemble average} of the power spectral density EA-PSD (lower curve) and the \textit{smoothed ensemble average} of the power spectral density SEA-PSD (upper curve, shifted in the vertical direction for clarity of presentation). The local maxima of the SEA-PSD [indicated by the ($\triangledown$) symbols] is used to estimate the mean value of the ripple length $l$ of each stage.}
\end{center}
\end{figure}
                                 
Figure \ref{fig:ripples_lje_stages} shows a portion of the surface for the four times that are indicated with symbols in Fig. \ref{fig:scaling_ripples_quasi_linear}. The left column shows the height $h^*_i$, which is the height minus its average value at each time $h^*_i=h_i-\langle h_i\rangle$. The right column shows the corresponding \textit{ensemble average} of the power spectral density EA-PSD computed with 30 realizations and $L=2048$. The PSD is represented in a logarithmic scale with arbitrary units and the horizontal axis corresponds to the wave number $k$. A moving window average over ten points is applied on the spectra and the resulting \textit{smoothed} ensemble average of the power spectral density SEA-PSD appears above the EA-PSD curve. The local maxima of the SEA-PSD (indicated by the $\triangledown$ symbols) can be used to estimate the mean value of the ripple length $l$ of each stage.

In the rough surface corresponding to $t=51$ $(\circ \ \mathrm{symbol})$ the integer values of the heights are visible. For $t=442\ (\square \  \mathrm{symbol})$ the ripples start to appear but their shape and length are highly irregular. More soft and larger ripples are found for times in the regime around to $t=11\,437\ (\bigtriangleup \  \mathrm{symbol})$ and the length estimated by the SEA-PSD method is approximately $l=23a$. It is observed that the ripple length is increasing with time due to the merging of ripples: small ripples are assimilated by contiguous larger ripples, which in turn develop a sinusoidal shape. For time $t=10^5 \ (\lozenge \  \mathrm{symbol})$ the ripple length is approximately $40a$. 

The increasing ripple length is a deviation from linear analysis predictions on the Kuramoto-Sivashinsky description. It is originated from nonlinear terms that can be examined in the derivation of the continuum equation for this discrete model as it is shown in Ref. \cite{mora_fe5:2005}. Analogous coarsening phenomena have been reported for experiments on ion sputtering \cite{habenicht:2002,rusponi:1998} and laser ablation \cite{georgescu:2004}, among others.

\section{Simulation of kerfs formation in LJE}\label{sec:laser_jet_action}

\begin{figure}
\begin{center}
\includegraphics[width=0.482\textwidth]{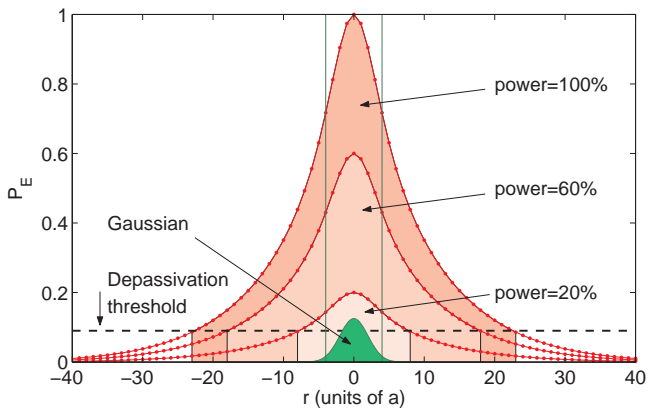}\caption{\label{fig:power_regime_pe}(Color online) Etching probability distribution $P_E$ associated with the estimation of the temperature field produced by a Gaussian beam with standard deviation $\sigma=2a$ for beam powers: 20, 60, and $100\,\%$. The etching probability $P_E$ determines how frequently the erosion and surface diffusion rules are applied, and is used to simulate the joint action of the moving laser beam and etching jet for different powers. The Gaussian distribution depicted in the figure has the same standard deviation of the laser beam. The vertical lines are located at $r=\pm2\sigma$.}
\end{center}
\end{figure}

In order to simulate the joint action of the laser beam and the etchant jet it is necessary to estimate the heat spreading resulting from the absorption of the laser beam. The intensity of the moving laser can be expressed by means of
\begin{equation}
G(r,t)=\frac{1}{\sigma\sqrt{2\pi}}e^{-(r-v_ft)^2/(2\sigma^2)},
\end{equation}
where $r$ is the spatial coordinate, $t$ the time, $\sigma$ corresponds to the standard deviation, and $v_f$ is the feed velocity. Given the thickness of the metallic sample ($\approx200\ \mu$m) and the relatively large dwell times of the laser beam, the back surface of the sample reaches approximately the same temperature as the front surface on which the radiation is incident. For this \textquotedblleft thermally thin approximation" (see Chapter 3 in Ref. \cite{ready:laser}), the computation of the temperature field requires numerical integration that will be reported in detail elsewhere. Regardless of variations related to the choice of the various thermal constants and parameters, a generic profile of the temperature field can be identified. The three curves shown in Fig. \ref{fig:power_regime_pe} are proportional to such a profile that can be described to be similar to a Gaussian shape within the laser spot while for larger distances it decays as $-log(r)$.

We define the \textit{etching probability distribution} $P_E$ of applying the erosion and surface diffusion rules to be directly proportional to the estimated temperature field. Introducing an additional factor between 0 and 1 to the $P_E$ distribution, it is possible to simulate laser beam powers between 0 and $100\,\%$. In order to simulate the fact that etching occurs only above a certain temperature, a depassivation threshold is introduced and $P_E$ is defined to be zero below it (see Fig. \ref{fig:power_regime_pe}). Therefore the etching probability distribution has a finite diameter $d$, which depends on power and depassivation threshold.  Figure \ref{fig:power_regime_pe} shows $P_E$ for three different power levels 20, 60 and $100\,\%$ and depassivation threshold of 0.09. The diameter and amplitude increase proportionally to the power. In order to compare the diameters $d$ of the three $P_E$ distributions (which are $16a$, $36a$, and $46a$, respectively) with the laser spot diameter used in the simulations ($d_{sim}\equiv4\sigma=8a$), the corresponding Gaussian distribution is also shown in the figure.

In the simulation the surface is kept fixed while the laser beam and the etching probability distribution move with velocity $v_f$. Within the etching front a combined action of erosion and surface diffusion rules occurs during a characteristic dwell time $d/v_f$. Depending upon $v_f$ and the power, the etching front will eventually develop instabilities giving rise to ripple structures. After the etching front leaves the region the surface has acquired a topography that, depending on model parameters, corresponds roughly to one of the stages analyzed in Fig. \ref{fig:scaling_ripples_quasi_linear}.

\begin{figure}
\begin{center}
\includegraphics[width=0.4\textwidth]{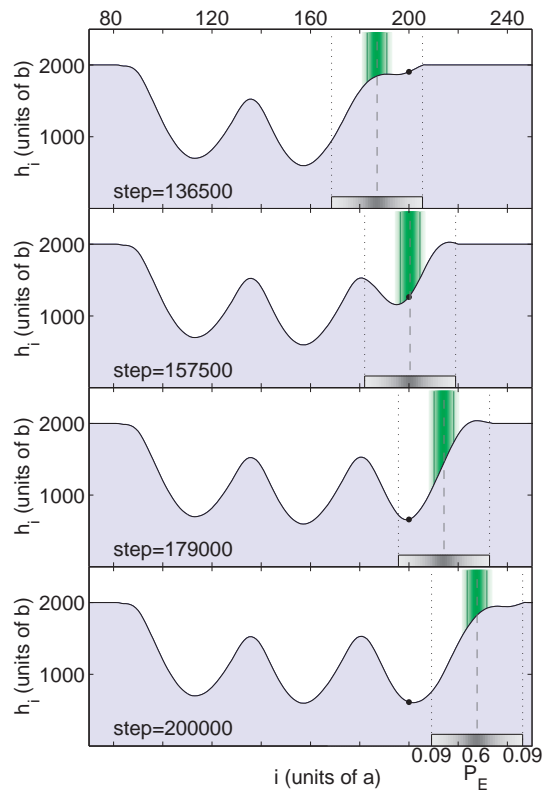}\caption{\label{fig:one_ripple_creation}(Color online) Four stages of the formation of a single ripple. The standard deviation of the moving Gaussian beam is $\sigma=2a$. Its center is represented by the vertical dashed line. The vertical continuous lines are located at $\pm2\sigma$ in a comoving system. The etching probability distribution $P_E$ corresponding to $60\,\%$ of power and depassivation threshold 0.09 shown in Fig. \ref{fig:power_regime_pe} is represented by the gray-scale bar at the bottom of each frame. A marker illustrates the evolution of a point on the surface at a fixed position $i=200$. The feed velocity is $2\times10^{-6} a$/step. Extended model parameters: $Y(\varphi)\equiv 1$, $f=0.045$, $J/(k_BT)=1$, $a=20$, $b=1$, $p_{\kappa ,min}=0.1$, $\kappa_{max}=0.0004$.}
\end{center}
\end{figure} 

Figure \ref{fig:one_ripple_creation} shows the creation process of a single ripple for a feed velocity $2\times10^{-6} a$/step and etching probability distribution $P_E$ corresponding to power $60\,\%$ shown in Fig. \ref{fig:power_regime_pe}. The extended model parameters are the same as used in Figs. \ref{fig:scaling_ripples_quasi_linear} and \ref{fig:ripples_lje_stages} of the previous section. Typically, a valley is created at the forefront of the moving $P_E$ from a small but growing depression on the surface. When the center of $P_E$ passes through a valley, the rate of erosion increases, and the valley grows as long as the rearmost part of the $P_E$ is acting. The peaks between the valleys are eroded at lower rate due to their negative curvature. In summary, the local and temporal action of the etching probability distribution works as an amplifier of small instabilities on the surface.

\begin{figure}
\begin{center}
\includegraphics[width=0.482\textwidth]{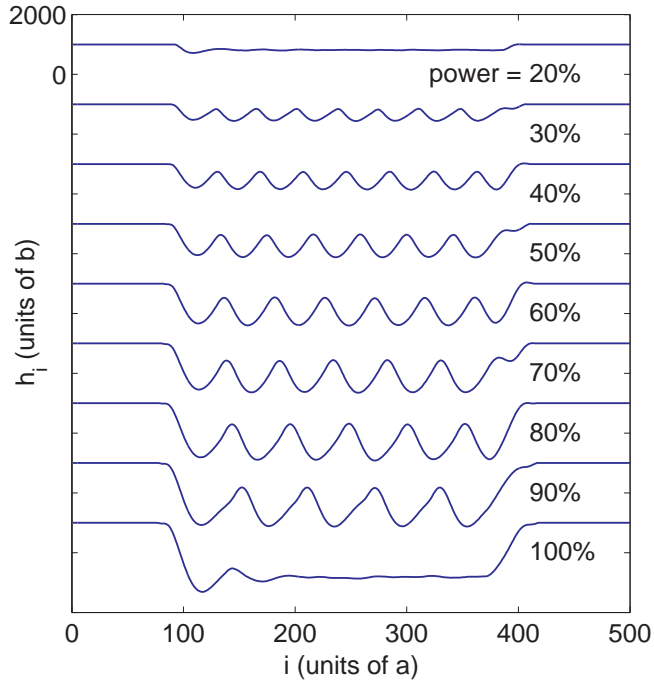}\caption{\label{fig:nine_powers}(Color online) Profiles of the kerfs for beam powers  $20,\,30,\ldots,100\,\%$ structured with the temperature related etching probability distributions $P_E$ shown in Fig. \ref{fig:power_regime_pe}. All kerfs are structured with the same feed velocity $2\times 10^{-6} a/$step. For power $20\,\%$ the kerf is shallow and rough while for powers from 30 to $90\,\%$ a ripple regime with increasing ripple length appears. For beam power $100\,\%$ the bottom of the kerf again becomes approximately flat with small roughness. The extended model parameters are the same as in the previous figure.}
\end{center}
\end{figure}

Figure \ref{fig:nine_powers} shows the profiles of the kerfs structured with etching probability distributions corresponding to powers ranging from $20$ to $100\,\%$ and constant feed velocity. For increasing power, the diameter $d$ of the corresponding $P_E$ distributions grows leading to an increase of the dwell time $d/v_f$. Together with the increase of the overall amplitude of the probability $P_E$, the resulting topographies correspond to later stages of the evolution shown in Fig. \ref{fig:scaling_ripples_quasi_linear}. For power $20\,\%$ the kerf is shallow and rough because its topography corresponds to early times of the evolution where the ripple regime is not yet reached. For powers ranging from 30 to $90\,\%$ regular ripples with increasing wavelength are obtained. For beam power $100\,\%$, after a few initial transient oscillations, the bottom of the kerf becomes approximately flat with small roughness.

\begin{figure}
\begin{center}
\includegraphics[width=0.482\textwidth]{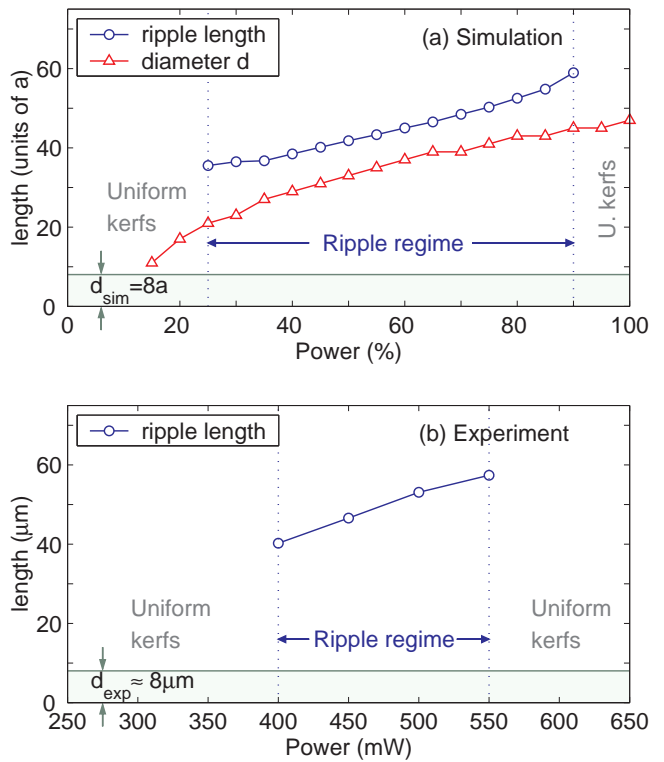}\caption{\label{fig:combo_length}(Color online) (a) Simulation. Increasing ripple length ($\circ$ symbols) for beam powers $25,\,30,\,35,\ldots,90\,\%$ corresponding to the same parameter set as in the two previous figures. For powers $15$ and $20\,\%$ the kerfs are shallow and uniform. For 95 and $100\,\%$ the kerfs become also uniform after a few transient oscillations. The diameters $d$ ($\vartriangle$ symbols) of the corresponding etching probability distributions $P_E$ increase with the power. (b) Experiment. Ripple length ($\circ$ symbols) of the kerfs shown in Fig. \ref{fig:power_series} corresponding to powers $400-550$ mW. The kerfs corresponding to powers $\lesssim350$ mW and $\gtrsim600$ mW are considered uniform. For both (a) and (b), the region between the two dotted vertical lines define approximately the power interval where periodic ripples appear.}
\end{center}                  
\end{figure}

The obtained ripples are very regular and periodic. An ensemble average over ten realizations with $16\,384$ points of the bottom of the kerf allow us to identify in the power spectral density a definite ripple length with negligible uncertainty. Figure \ref{fig:combo_length}(a) shows that the ripple length grows almost linearly with the power ($\circ$ symbols). The diameter $d$ of the etching probability distributions (depicted with the $\vartriangle$ symbols) also grows with the power. In order to compare with the ripple lengths, the diameter of the laser spot $d_{sim}=8a$ used in the simulation is indicated in the figure with the horizontal line. For powers $\lesssim20\,\%$ and for powers $\gtrsim95\,\%$ the obtained kerfs are uniform.

In order to compare our model with the experiment, Fig. \ref{fig:combo_length}(b) shows the ripple lengths corresponding to powers 400, 450, 500, and 550 mW of Fig. \ref{fig:power_series}. The ripple length increases proportionally with the laser power. The diameter of the laser spot $d_{exp}\approx8\ \mu$m used in the experiment is indicated in the figure with the horizontal line. Of course, four points are not enough to draw conclusions about this dependency. We will explore further this correspondence when more experimental data become available. Summarizing, the simulations show a qualitative agreement of our hypotheses with the experiment: the etching front covers a region wider than the laser spot and the resulting characteristic ripple length is proportional to the diameter of the active etching zone.

\section{Conclusions}\label{sec:conclusions}

The discrete extended model applied to the simulation of formation of kerfs reproduces qualitatively some of the main characteristics of microstructures produced by the LJE experiment. The basic mechanism based on the competition of curvature dependent erosion and surface diffusion generates a ripple regime, which appears between roughening regimes at earlier and later stages of the surface evolution. Compared with the original CMTHS model, the extended model produces longer time intervals where ripples occur, which is more appropriate for the simulation of the moving etching source of the LJE experiment. This is accomplished by applying the erosion and diffusion rules within a comoving etching probability distribution that is proportional to an estimated temperature field produced by the laser spot. Taking advantage of the probabilistic nature of the Monte Carlo method, it is possible to simulate the laser power distribution, laser spot diameter and feed velocity. In analogy with the experiments, variations of the laser power reveal a regime with an unstable etching front and in consequence, regular ripples in the bottom of the kerfs. In addition, the ripple length is growing with increasing power. Outside of this interval of power values, the kerfs are uniform with small roughness at the bottom.  

This simple phenomenological model does not pretend to simulate all details of the complex behavior of the process. Instead, it can be considered as a guide to propose experiments that reveal more relevant features of the ripple formation and the structuring process in general. The next step is to incorporate the slope dependency of the absorption of the laser polarized light into the model. In addition, including the dynamics of etchant concentration in the Nernst diffusion layer together with an additional heat source due to the exothermic reactions could allow us to understand the conditions for the onset of the thermal runaway responsible of ripple formation. For a better comparison with the experimental results a further obvious step is the generalization of the extended model to 2+1 dimensions.

\begin{acknowledgments}
We gratefully acknowledge Simeon Metev and Andreas Stephen for the support at the BIAS Institute. We thank Rudolf Friedrich, Stefan Linz, Hans J. Herrmann, Rudolf Hilfer, Alexei Kouzmitchev, Bernd Lehle, Volker Schatz, and Ferenc Kun for helpful discussions and suggestions. This work was funded by Volkswagen-Stiftung Grant I/77315.
\end{acknowledgments}

\bibliographystyle{apsrev}
%\bibliography{/disk2/data/thesis/references/biblio}

\end{document}